\documentclass[twocolumn,showpacs,aps,prl,superscriptaddress]{revtex4}
\usepackage{graphicx}
\usepackage{lineno}
\usepackage{dcolumn}
\usepackage{amsmath}
\usepackage{epsfig}
\usepackage{multirow}

\newcommand{\BaBarYear}       {05}
\newcommand{\BaBarNumber}     {035}
\newcommand{\SLACPubNumber} {11334}

\RequirePackage{xspace}

\usepackage{relsize}
\def\babar{\mbox{\slshape B\kern-0.1em{\smaller A}\kern-0.1em
    B\kern-0.1em{\smaller A\kern-0.2em R}}}

\def\epem       {\ensuremath{e^+e^-}\xspace}

\def\qqbar {\ensuremath{q\overline q}\xspace}

\def\pip   {\ensuremath{\pi^+}\xspace}
\def\pim   {\ensuremath{\pi^-}\xspace}

\def\pipm  {\ensuremath{\pi^\pm}\xspace}

\def\Kbar  {\kern 0.2em\overline{\kern -0.2em K}{}\xspace}

\def\Kz    {\ensuremath{K^0}\xspace}
\def\Kzb   {\ensuremath{\Kbar^0}\xspace}
\def\KzKzb {\ensuremath{\Kz \kern -0.16em \Kzb}\xspace}
\def\Kp    {\ensuremath{K^+}\xspace}
\def\Km    {\ensuremath{K^-}\xspace}
\def\Kpm   {\ensuremath{K^\pm}\xspace}

\def\KpKm  {\ensuremath{\Kp \kern -0.16em \Km}\xspace}
\def\KS    {\ensuremath{K^0_{\scriptscriptstyle S}}\xspace} 
\def\KL    {\ensuremath{K^0_{\scriptscriptstyle L}}\xspace}

\def\Dbar    {\kern 0.2em\overline{\kern -0.2em D}{}\xspace}

\def\Dz      {\ensuremath{D^0}\xspace}
\def\Dzb     {\ensuremath{\Dbar^0}\xspace}
\def\DzDzb   {\ensuremath{\Dz {\kern -0.16em \Dzb}}\xspace}
\def\Dp      {\ensuremath{D^+}\xspace}
\def\Dm      {\ensuremath{D^-}\xspace}

\def\DpDm    {\ensuremath{\Dp {\kern -0.16em \Dm}}\xspace}

\def\Bbar    {\kern 0.18em\overline{\kern -0.18em B}{}\xspace}

\def\BB      {\ensuremath{B\Bbar}\xspace} 
\def\Bz      {\ensuremath{B^0}\xspace}
\def\Bzb     {\ensuremath{\Bbar^0}\xspace}
\def\BzBzb   {\ensuremath{\Bz {\kern -0.16em \Bzb}}\xspace}
\def\Bu      {\ensuremath{B^+}\xspace}
\def\Bub     {\ensuremath{B^-}\xspace}
\def\Bp      {\ensuremath{\Bu}\xspace}
\def\Bm      {\ensuremath{\Bub}\xspace}

\def\BpBm    {\ensuremath{\Bu {\kern -0.16em \Bub}}\xspace}

\def\BorBbar    {\kern 0.18em\optbar{\kern -0.18em B}{}\xspace}
\def\DorDbar    {\kern 0.18em\optbar{\kern -0.18em D}{}\xspace}
\def\KorKbar    {\kern 0.18em\optbar{\kern -0.18em K}{}\xspace}

\mathchardef\Upsilon="7107
\def\Y#1S{\ensuremath{\Upsilon{(#1S)}}\xspace}

\def\FourS {\Y4S}

\mathchardef\Deltares="7101
\mathchardef\Xi="7104
\mathchardef\Lambda="7103
\mathchardef\Sigma="7106
\mathchardef\Omega="710A

\def\Deltabar{\kern 0.25em\overline{\kern -0.25em \Deltares}{}\xspace}
\def\Lbar{\kern 0.2em\overline{\kern -0.2em\Lambda\kern 0.05em}\kern-0.05em{}\xspace}
\def\Sigbar{\kern 0.2em\overline{\kern -0.2em \Sigma}{}\xspace}
\def\Xibar{\kern 0.2em\overline{\kern -0.2em \Xi}{}\xspace}
\def\Obar{\kern 0.2em\overline{\kern -0.2em \Omega}{}\xspace}
\def\Nbar{\kern 0.2em\overline{\kern -0.2em N}{}\xspace}
\def\Xb{\kern 0.2em\overline{\kern -0.2em X}{}\xspace}

\def\BR         {{\ensuremath{\cal B}\xspace}}

\def\upsbzbz {\ensuremath{\FourS \to \BzBzb}\xspace}

\def\mes        {\mbox{$m_{\rm ES}$}\xspace}

\newcommand{\tev}{\ensuremath{\mathrm{\,Te\kern -0.1em V}}\xspace}
\newcommand{\gev}{\ensuremath{\mathrm{\,Ge\kern -0.1em V}}\xspace}
\newcommand{\mev}{\ensuremath{\mathrm{\,Me\kern -0.1em V}}\xspace}
\newcommand{\kev}{\ensuremath{\mathrm{\,ke\kern -0.1em V}}\xspace}
\newcommand{\ev}{\ensuremath{\mathrm{\,e\kern -0.1em V}}\xspace}
\newcommand{\gevc}{\ensuremath{{\mathrm{\,Ge\kern -0.1em V\!/}c}}\xspace}
\newcommand{\mevc}{\ensuremath{{\mathrm{\,Me\kern -0.1em V\!/}c}}\xspace}
\newcommand{\gevcc}{\ensuremath{{\mathrm{\,Ge\kern -0.1em V\!/}c^2}}\xspace}
\newcommand{\mevcc}{\ensuremath{{\mathrm{\,Me\kern -0.1em V\!/}c^2}}\xspace}

\def\mus  {\ensuremath{\rm \,\mus}\xspace}

\def\mus        {\ensuremath{\,\mu{\rm s}}\xspace}    

\def\to                 {\ensuremath{\rightarrow}\xspace}

\def\pep2{PEP-II}

\def\gsim{{~\raise.15em\hbox{$>$}\kern-.85em
          \lower.35em\hbox{$\sim$}~}\xspace}
\def\lsim{{~\raise.15em\hbox{$<$}\kern-.85em
          \lower.35em\hbox{$\sim$}~}\xspace}

\def\eps{\varepsilon\xspace}

\def\CP                {\ensuremath{C\!P}\xspace}

\xspace

\newcommand{\epjBase}        {Eur.\ Phys.\ Jour.\xspace}
\newcommand{\jprlBase}       {Phys.\ Rev.\ Lett.\xspace}
\newcommand{\jprBase}        {Phys.\ Rev.\xspace}
\newcommand{\jplBase}        {Phys.\ Lett.\xspace}
\newcommand{\nimBaseA}       {Nucl.\ Instr.\ Methods Phys.\ Res., Sect.\ A\xspace}

\newcommand{\npBase}         {Nucl.\ Phys.\xspace}

\newcommand{\epjc}      [1]  {\epjBase\ C~{\bf #1}}

\newcommand{\npb}       [1]  {\npBase\ B~{\bf #1}}

\newcommand{\npbps}     [1]  {{Nucl.\ Phys.\ B~Proc.\ Suppl.\ {\bf #1}}}

\newcommand{\plb}       [1]  {\jplBase\ B~{\bf #1}}

\newcommand{\jprl}      [1]  {\jprlBase\ {\bf #1}}
\newcommand{\jprd}      [1]  {\jprBase\ D~{\bf #1}}

\newcommand{\progtp}    [1]  {{Prog.\ Theor.\ Phys.\ {\bf #1}}}

\def\geant      {\mbox{\tt GEANT}\xspace}

\def\jetset74   {\mbox{\tt Jetset \hspace{-0.5em}7.\hspace{-0.2em}4}\xspace}

\def\de {\ensuremath{\Delta E}}

\RequirePackage{xspace}

\def\figurebox#1#2#3{%
    \def\arg{#3}%
    \ifx\arg\empty
    {\hfill\vbox{\hsize#2\hrule\hbox to #2{\vrule\hfill\vbox to #1{\hsize#2\vfill}\vrule}\hrule}\hfill}%
    \else
    {\hfill\epsfbox{#3}\hfill}%
    \fi}

\long\def\inst#1{\par\nobreak\kern 4pt\nobreak
    {\it #1}\par\vskip 10pt plus 3pt minus 3pt}

\begin{document}

\begin{flushleft}
\babar-PUB-\BaBarYear/\BaBarNumber\\
SLAC-PUB-\SLACPubNumber
\end{flushleft}

\title{
{
\Large \bf \boldmath
Evidence for $\Bu\to\Kzb\Kp$ and $\Bz\to\Kz\Kzb$, and Measurement of
the Branching Fraction and Search for Direct \CP\ Violation in 
$\Bu\to\Kz\pip$}}

%
\author{B.~Aubert}
\author{R.~Barate}
\author{D.~Boutigny}
\author{F.~Couderc}
\author{Y.~Karyotakis}
\author{J.~P.~Lees}
\author{V.~Poireau}
\author{V.~Tisserand}
\author{A.~Zghiche}
\affiliation{Laboratoire de Physique des Particules, F-74941 Annecy-le-Vieux, France }
\author{E.~Grauges}
\affiliation{IFAE, Universitat Autonoma de Barcelona, E-08193 Bellaterra, Barcelona, Spain }
% ...
\collaboration{The \babar\ Collaboration}
\noaffiliation

\begin{abstract}
We present evidence for the $b\to d$ penguin-dominated decays 
$\Bu\to\Kzb\Kp$ and $\Bz\to\Kz\Kzb$ with significances of 
$3.5$ and $4.5$ standard deviations, respectively.  The results 
are based on a sample of $227$ million $\Y4S\to\BB$ decays collected with 
the \babar\ detector at the \pep2\ asymmetric-energy $\epem$ collider 
at SLAC.  We measure the branching fractions 
$\BR(\Bu\to\Kzb\Kp) = (1.5\pm 0.5\pm 0.1)\times 10^{-6}\,(< 2.4\times 10^{-6})$
and $\BR(\Bz\to\Kz\Kzb) = (1.19^{+0.40}_{-0.35}\pm 0.13)\times 10^{-6}$, where
the uncertainties are statistical and systematic, respectively, and the upper
limit on the branching fraction for $\Kzb\Kp$ is at the $90\%$ confidence
level.  We also present improved measurements of the charge-averaged 
branching fraction $\BR(\Bu\to\Kz\pip) = (26.0\pm 1.3\pm 1.0)\times 10^{-6}$
and \CP-violating charge asymmetry 
${\cal A}_{\CP}\,(\Kz\pip) = -0.09\pm 0.05\pm 0.01$, where the uncertainties
are statistical and systematic, respectively.
\end{abstract}


\maketitle

Flavor-changing neutral currents are forbidden at first order in the standard
model, but can proceed through weak interactions that are described by 
one-loop ``penguin'' diagrams.  Such decay processes were first established in 
the $B$ system more than a decade ago through observation of the radiative 
decay $B\to K^*\gamma$~\cite{CleoKG}, which is dominated by the $b\to s\gamma$ 
electromagnetic-penguin amplitude.  Recently, the analogous gluonic-penguin 
process $b\to sg~(g\to s\bar{s})$ has been used extensively to test the
standard model predictions for the \CP-violating asymmetry amplitudes of decay 
modes such as $\Bz\to\phi\KS$~\cite{phiks}.  To date, no direct evidence has 
been found for decays dominated by the corresponding $b\to dg$ transition, 
whose amplitude is suppressed relative to that for the $b\to sg$ process by 
the small ratio $V_{\rm td}/V_{\rm ts}$ involving elements of the 
Cabibbo-Kobayashi-Maskawa quark-mixing matrix~\cite{CKM}.  Such decays could 
play an important complementary role in the search for new physics in the $B$ 
system.

In this Letter, we report evidence for the decays $\Bu\to\Kzb\Kp$ and 
$\Bz\to\Kz\Kzb$, which are expected to be dominated by the 
$b\to dg~(g\to s\bar{s})$ penguin diagram, and an updated measurement of the 
branching fraction and direct \CP-violating charge asymmetry for 
$\Bu\to\Kz\pip$ (the use of charge conjugate modes is implied throughout this 
paper unless otherwise stated).  Our previous search for the $K\Kzb$ modes 
yielded branching-fraction upper limits at the level of 
$2\times 10^{-6}$~\cite{KsX04}, which are consistent with recent theoretical 
estimates based on perturbative calculations~\cite{theory}, as well as the 
lower bounds implied by $SU(3)$ symmetry~\cite{FleischerRecksiegel}.  

Once the decay $\Bz\to\Kz\Kzb$ has been established, a measurement of its 
time-dependent \CP-violating asymmetry (through the technique described in
Ref.~\cite{BaBarPi0Ks}) could provide important constraints on physics beyond 
the standard model.  Assuming top-quark dominance in the penguin loop, the 
asymmetry is expected to vanish in the standard model~\cite{LondonQuinn}, 
while contributions from supersymmetric particles could be 
significant~\cite{Giri}.  Although soft rescattering effects could 
weaken the sensitivity to new physics in this mode~\cite{Fleischer94}, the 
ratio of decay rates for $\Bu\to\Kzb\Kp,\,\Kz\pip$ can be used to constrain 
the relative size of such effects~\cite{Fleischer99}.

Recent measurements of the partial-rate asymmetry in $\Bz\to\Kp\pim$ decays
by the \babar~\cite{KpiBaBar} and Belle~\cite{KpiBelle} experiments have 
established direct \CP\ violation in the $B$ system.  In this Letter, we 
search for direct \CP\ violation in the decays $\Bu\to\Kz\pip,\,\Kzb\Kp$ 
through measurement of the charge asymmetry
\begin{equation*}
{\cal A}_{\CP} = \frac{\Gamma(\Bub\to f^-) - \Gamma(\Bu\to f^+)}
{\Gamma(\Bub\to f^-) + \Gamma(\Bu\to f^+)},
\end{equation*}
where $f^{\pm} = \KS\pipm,~\KS\Kpm$.
The decay $\Bub\to\Kzb\pim$ is dominated by the $b\to s$ penguin process 
and, neglecting rescattering effects~\cite{Fleischer99}, is expected to yield 
${\cal A}_{\CP}\sim 1\%$~\cite{theory,Ciuchini01}.  Observation 
of a significant charge asymmetry could therefore indicate new 
physics entering the penguin loop~\cite{Falk98}.  The decay rate and charge 
asymmetry in $\KS\pip$ can also be used to constrain the angle $\gamma$ of
the unitarity triangle~\cite{gamma}.

The data sample used in this analysis contains $(226.6\pm 2.5)\times 10^6$
$\Y4S\to\BB$ decays collected by the \babar\ detector~\cite{babar} at the
SLAC \pep2\ asymmetric-energy $\epem$ collider.  The primary detector
elements used in this analysis are a charged-particle tracking system 
consisting of a five-layer silicon vertex tracker (SVT) and a 40-layer drift 
chamber (DCH) surrounded by a $1.5$-T solenoidal magnet, and a dedicated 
particle-identification system consisting of a detector of internally 
reflected Cherenkov light (DIRC).

We identify two separate event samples corresponding to the decay topologies
$\Bz\to \KS\KS$ and $\Bu\to \KS h^+$, where $h^{\pm}$ is either a pion or a 
kaon.  Neutral kaons are reconstructed in the mode $\KS\to\pip\pim$ by 
combining pairs of oppositely charged tracks originating from a common decay 
point and having a $\pip\pim$ invariant mass within $11.2\mevcc$ of the 
nominal $\KS$ mass~\cite{PDG2004}.  To reduce combinatorial background, we 
require the measured proper decay time of the $\KS$ to be greater than five 
times its uncertainty.  Candidate $h^+$ tracks are assigned the 
pion mass and are required to originate from the interaction region and to have 
an associated Cherenkov angle $(\theta_c$) measurement with at least six signal 
photons detected in the DIRC.  To reduce backgrounds from protons and leptons, 
we require $\theta_c$ to be within $4$ standard deviations ($\sigma$) of the 
expectation for either the pion or kaon particle hypothesis.  
The $\Bz$ sample is formed by combining pairs of \KS\ candidates, 
while the $\Bu$ sample is formed by combining \KS\ and $h^+$ candidates.

We exploit the unique kinematic and topological features of charmless
two-body $B$ decays to suppress the dominant background arising from
the process $\epem\to\qqbar\,(q=u,d,s,c)$.  For each $\Bz$ candidate,
we require the difference $\de$ between its reconstructed 
center-of-mass (CM) energy and the beam energy $(\sqrt s/2)$ to
be less than $100\mev$.  For $\Bu$ candidates, we require
$-115 < \de < 75\mev$, where the lower limit accounts for an average
shift in $\de$ of $-45\mev$ in the $\Kzb\Kp$ mode due to the assignment of 
the pion mass to the $\Kp$.  We also define a beam-energy substituted mass 
$\mes \equiv \sqrt{(s/2 + {\mathbf {p}}_i\cdot {\mathbf {p}}_B)^2/E_i^2 - 
{\mathbf {p}}_B^2}$, where the $B$-candidate momentum ${\mathbf {p}}_B$ 
and the four-momentum of the initial $\epem$ state 
$(E_i, {\mathbf {p}}_i)$ are calculated in the laboratory frame.  We require
$5.20 < \mes < 5.29\gevcc$ for $B$ candidates in both samples.  To suppress 
the jet-like $\qqbar$ background, we calculate the CM angle $\theta^*_S$ 
between the sphericity axis of the $B$ candidate and the sphericity axis of 
the remaining charged and neutral particles in the event, and require 
$\left | \cos(\theta^*_S) \right | < 0.8$.

\renewcommand{\multirowsetup}{\centering}
\newlength{\LL}\settowidth{\LL}{$8047$}
\begin{table*}[!htb]
\begin{center}
\caption{Summary of results for the total detection efficiencies 
$\eps$, fitted signal yields $n$, signal-yield significances $s$ (including
systematic uncertainty), charge-averaged branching fractions $\BR$, and 
charge asymmetries ${\cal A}_{CP}$ (including $90\%$ confidence intervals).  
The efficiencies include the branching fraction for
$\KS\to\pip\pim$ and the probability of $50\%$ 
for $\Kz\Kzb\to \KS\KS$.  Branching fractions are calculated 
assuming equal rates for $\upsbzbz$ and $\Bp\Bm$~\cite{BBratio}.  
For $\Kzb\Kp$, we give both the central value of the branching fraction
and, in parentheses, the $90\%$ confidence-level (CL) upper limit.}
\label{tab:summary}
\begin{ruledtabular}
\begin{tabular}{lcccccc}
Mode  & $\eps$ (\%) & $n$ & $s~(\sigma)$ & \BR~($10^{-6}$) & ${\cal A}_{\CP}$ &
${\cal A}_{\CP}\,(90\% {\rm CL})$\\
\hline \\[-3mm]
$\Bu\to\Kz\pip$  & $12.6 \pm 0.3$ & $744^{+37\;+21}_{-36\;-17}$  &
           & $26.0 \pm 1.3 \pm 1.0$ & $-0.09\pm 0.05\pm 0.01$ & $[-0.16,-0.02]$\\[2mm]
$\Bu\to\Kzb\Kp$  & $12.5 \pm 0.3$ & $41^{+15\;+3}_{-13\;-2}$ &
            $3.5$ & $1.5\pm 0.5\pm 0.1\,(<2.4)$ & $0.15\pm 0.33\pm 0.03$ & $[-0.43,0.68]$ \\[2mm]
$\Bz\to\KzKzb$   & $8.5 \pm 0.6$ & $23^{+8}_{-7}\pm 2$ &
            $4.5$ & $1.19^{+0.40}_{-0.35}\pm 0.13$ & \\
\end{tabular}
\end{ruledtabular}
\end{center}
\end{table*}

After applying all of the above requirements, we find $1939$ $(20441)$
candidates in the $\Bz\,(\Bu)$ samples, respectively.  The fraction of 
events containing more than one $B$ candidate is negligible ($<0.5\%$).  
The total detection efficiencies are given in Table~\ref{tab:summary} and
include the branching fraction for $\KS\to\pip\pim$~\cite{PDG2004} 
and a probability of $50\%$ for $\Kz\Kzb\to \KS\KS$~\cite{KsKl}.
We use data and simulated Monte Carlo samples~\cite{geant} to verify that 
backgrounds from other $B$ decays are negligible.  The selected samples are 
therefore assumed to be composed of signal $B$ decays and background candidates 
arising from random combinations of tracks and $\KS$ mesons in $\qqbar$ events.  

To determine signal yields in each sample, we apply separate unbinned
maximum-likelihood fits incorporating discriminating variables that
account for differences between $\BB$ and $\qqbar$ events.  In addition 
to the kinematic variables $\mes$ and $\de$, we include a Fisher 
discriminant ${\cal F}$~\cite{pipi2002} defined as an optimized linear 
combination of the event-shape variables $\sum_i p_i^*$ and 
$\sum_i p_i^*\cos^2(\theta_i^*)$, where $p_i^*$ is the CM momentum of 
particle $i$, $\theta_i^*$ is the CM angle between the momentum of particle 
$i$ and the $B$-candidate thrust axis, and the sum is over all particles in 
the event excluding the $B$ daughters.  

The likelihood function to be maximized is defined as
\begin{equation*}
{\cal L} = \exp{\left (-\sum_{i}n_i \right )}
\prod_{j=1}^N\left[\sum_i n_i{\cal P}_i\right],
\end{equation*}
where $n_i$ and ${\cal P}_i$ are the yield and probability density
function (PDF) for each component $i$ in the fit, and $N$ is the total
number of events in the sample.  For the $\Bz$ sample there
are only two components (signal and background), and the total PDF is
calculated as the product of the individual PDFs for $\mes$, $\de$,
and ${\cal F}$.  We combine $\Bu$ and $\Bub$ candidates in a single fit 
and include the PDF for $\theta_c$ to determine separate yields and charge 
asymmetries for the two signal components, $\KS\pi$ and $\KS K$, and two 
corresponding background components.  For both signal and background, the 
$\KS h^{\pm}$ yields are parameterized as 
$n_{\pm} = n(1 \mp {\cal A}_{\CP})/2$; we fit directly for the total yield 
$n$ and the charge asymmetry ${\cal A}_{\CP}$.

The parameterizations of the PDFs are determined from data wherever
possible.  For the $\Bu$ sample, the large signal $\KS\pip$
component allows for an accurate determination of the peak positions
for $\mes$ and $\de$, as well as the parameters describing the shape 
of the PDF for ${\cal F}$.  We therefore allow these parameters
to vary freely in the fit.  The remaining shape parameters describing
$\mes$ and $\de$ are determined from simulated Monte Carlo 
samples and are fixed in the fit.  Except for the mean
value of $\de$, which is shifted by our use of the pion mass hypothesis for
the $h^+$ candidate, we use the $\KS\pip$ parameters to describe signal
$\KS\Kp$ decays.  The parameters describing the background PDFs in $\mes$ and 
${\cal F}$ are allowed to vary freely in the fit, while the $\de$ parameters 
are determined in the signal-free region of $\mes$ ($5.20< \mes < 5.26\gevcc$) 
and fixed in the fit.  For both signal and background, the $\theta_c$ PDFs are 
obtained from a sample of $D^{*+}\to \Dz\pip\,(\Dz\to\Km\pip)$ decays 
reconstructed in data, as described in Ref.~\cite{KpiBaBar}.
For the $\Bz$ sample, all shape parameters describing the signal PDFs 
are fixed to the values determined from Monte Carlo simulation, while the peak 
positions for $\mes$ and $\de$ are derived from the results of the fit to the 
$\Bu$ sample.  We allow the background ${\cal F}$ shape parameters to vary 
freely, while the PDF parameters for $\mes$ and $\de$ are fixed to the values 
determined from data in the signal-free regions 
$100 < \left |\de\right | < 300\mev$ (for $\mes$) and 
$5.20<\mes <5.26\gevcc$ (for $\de$).  

Several cross-checks were performed to validate the fitting 
technique before data in the signal region were examined.  We confirmed 
the internal self-consistency of the fitting algorithm by 
generating and fitting a large set of pseudo-experiments where
signal and background events were generated randomly from the PDFs
with yields corresponding to the expected values based on our
previous analysis of these modes~\cite{KsX04}.  The fitted
signal yields for all three modes were unbiased.  Correlations 
among the discriminating variables in background data events are 
found to be negligible.  To check for
residual correlations between the discriminating variables in
signal events, we performed a second test for the $\KS\KS$ mode
where simulated Monte Carlo samples of signal events were mixed 
with background events generated directly from the PDFs.  We
observed an average bias corresponding to approximately one event 
and include this effect in the systematic uncertainty on the
fitted $\KS\KS$ yield.  Potential $\KS\pi\to \KS K$ cross-feed
was evaluated by fitting large samples of simulated
Monte Carlo signal events.  The resulting small $(<0.5\%)$
biases are included in the systematic uncertainty on the fitted
yields.

The fit results supersede our previous measurements of these
quantities and are summarized in Table~\ref{tab:summary}.
The signal yields for $\Bu\to\KS\Kp$ and $\Bz\to\KS\KS$ correspond 
to significances of $3.5\sigma$ and $4.5\sigma$ (including
systematic uncertainties~\cite{significance}), respectively, and are 
consistent with our previous results~\cite{KsX04}, as well as with 
the results of other experiments~\cite{BelleAndCleo}.  The signal yield for  
$\Bu\to\KS\pip$ is somewhat higher than expected from our previous 
result.  A re-analysis of the first $88$ million $\BB$ events 
yields $285\pm 21$ $\KS\pip$ signal events, compared with $255\pm 20$ 
reported in Ref.~\cite{KsX04}.  Approximately half of this difference is 
due to reprocessing of the data with improved calibration constants.  The 
remaining difference is due to improved knowledge of the PDF parameters, which 
were the largest source of systematic uncertainty for the previous result.  We 
find $459\pm 29$ events in the remaining $139$ million $\BB$ events, which is 
consistent with the signal yield obtained in the first part of the sample.

For the $\KS\Kp$ mode, we compute an upper limit on the signal yield
as the value of $n_0$ for which 
$\int_0^{n_0}{\cal L}_{\rm max}{\rm d}n/\int_0^{\infty}{\cal L}_{\rm max}{\rm d}n = 0.9$,
where ${\cal L}_{\rm max}$ is the likelihood as a function of $n$, maximized
with respect to the remaining free parameters.  The corresponding 
branching-fraction upper limit is calculated by increasing $n_0$
and reducing the efficiency by their respective systematic 
uncertainties.

\begin{figure}[!tbp]
\begin{center}
\includegraphics[width=0.50\linewidth]{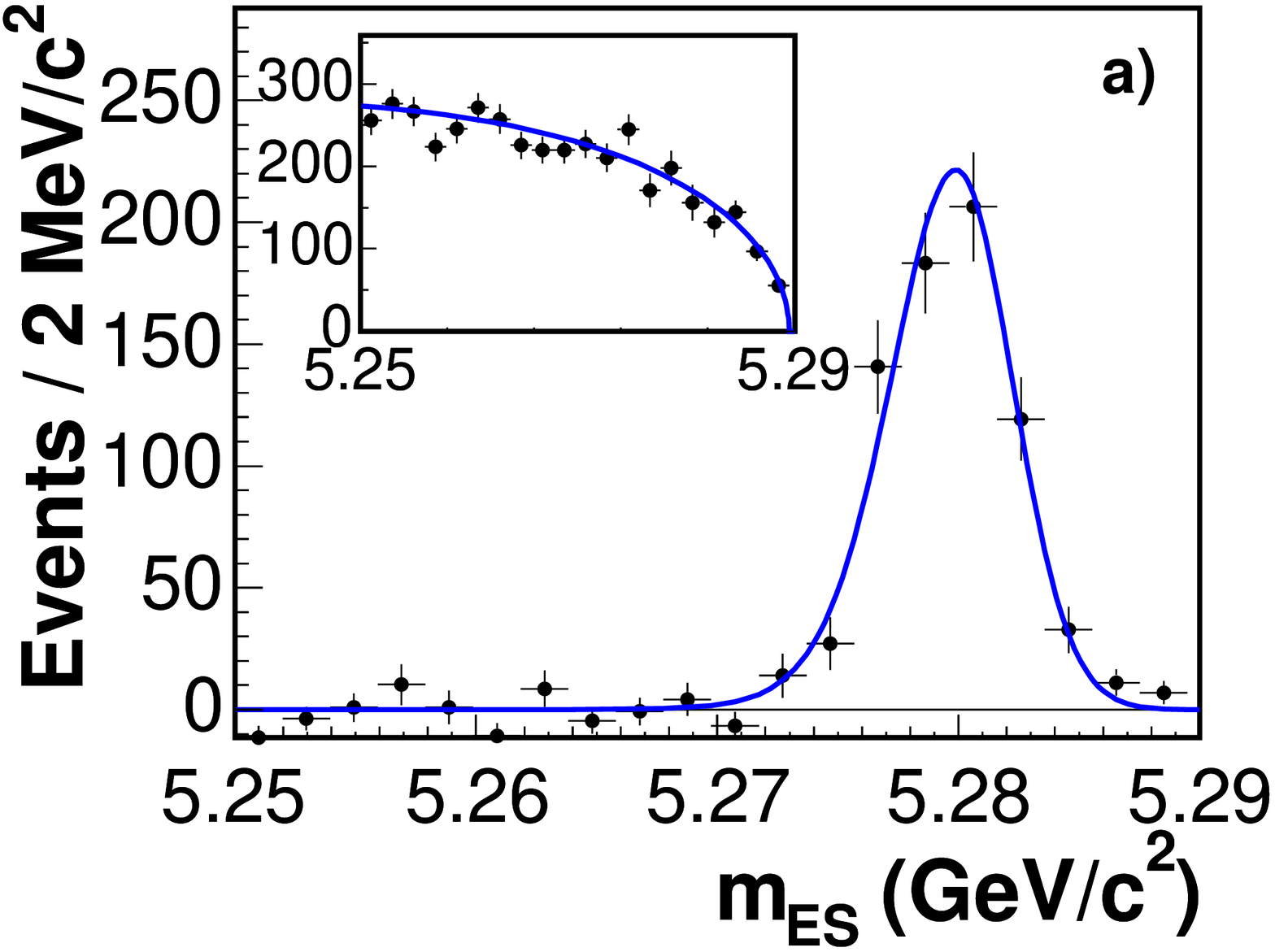}\includegraphics[width=0.50\linewidth]{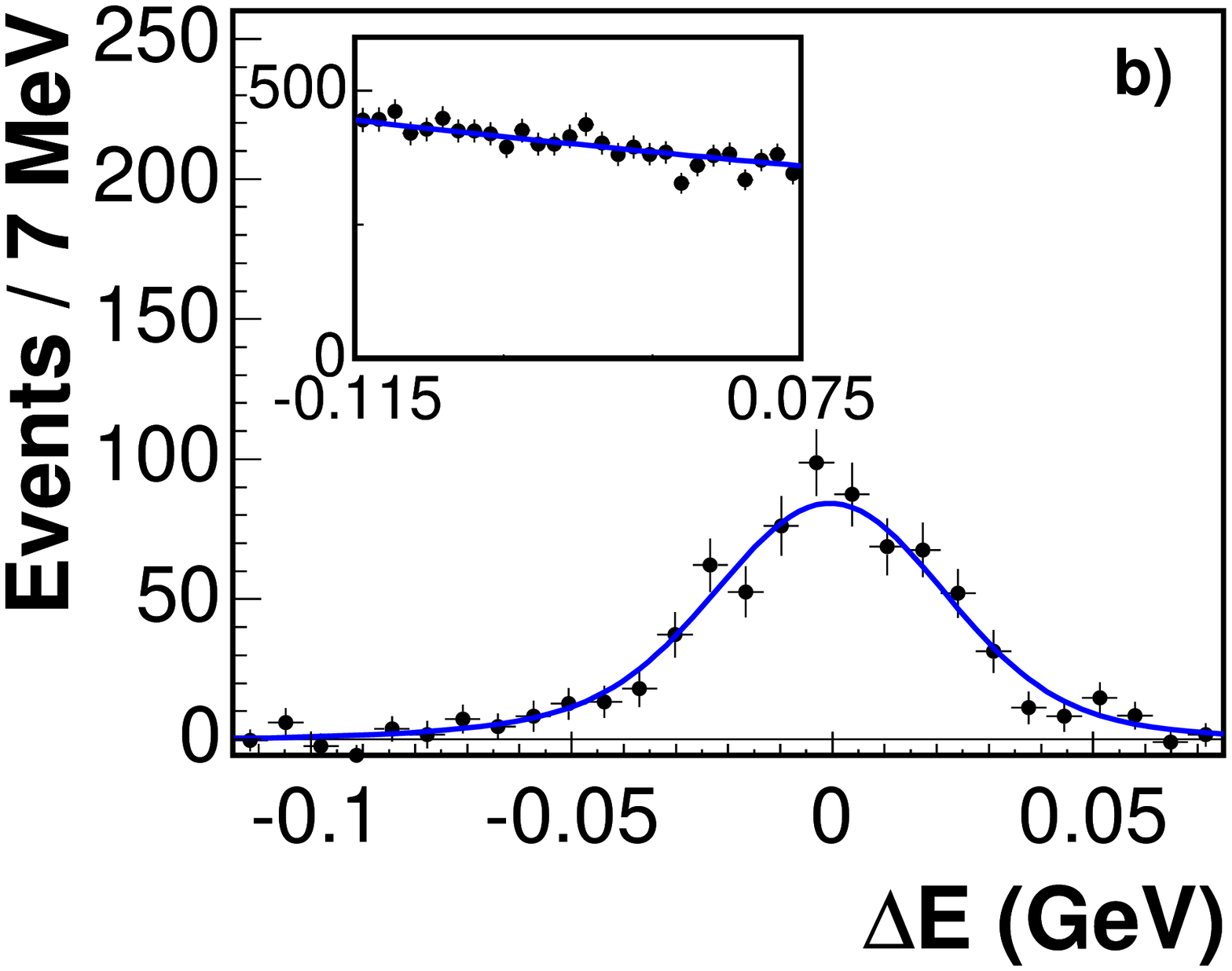}
\vskip1mm
\includegraphics[width=0.50\linewidth]{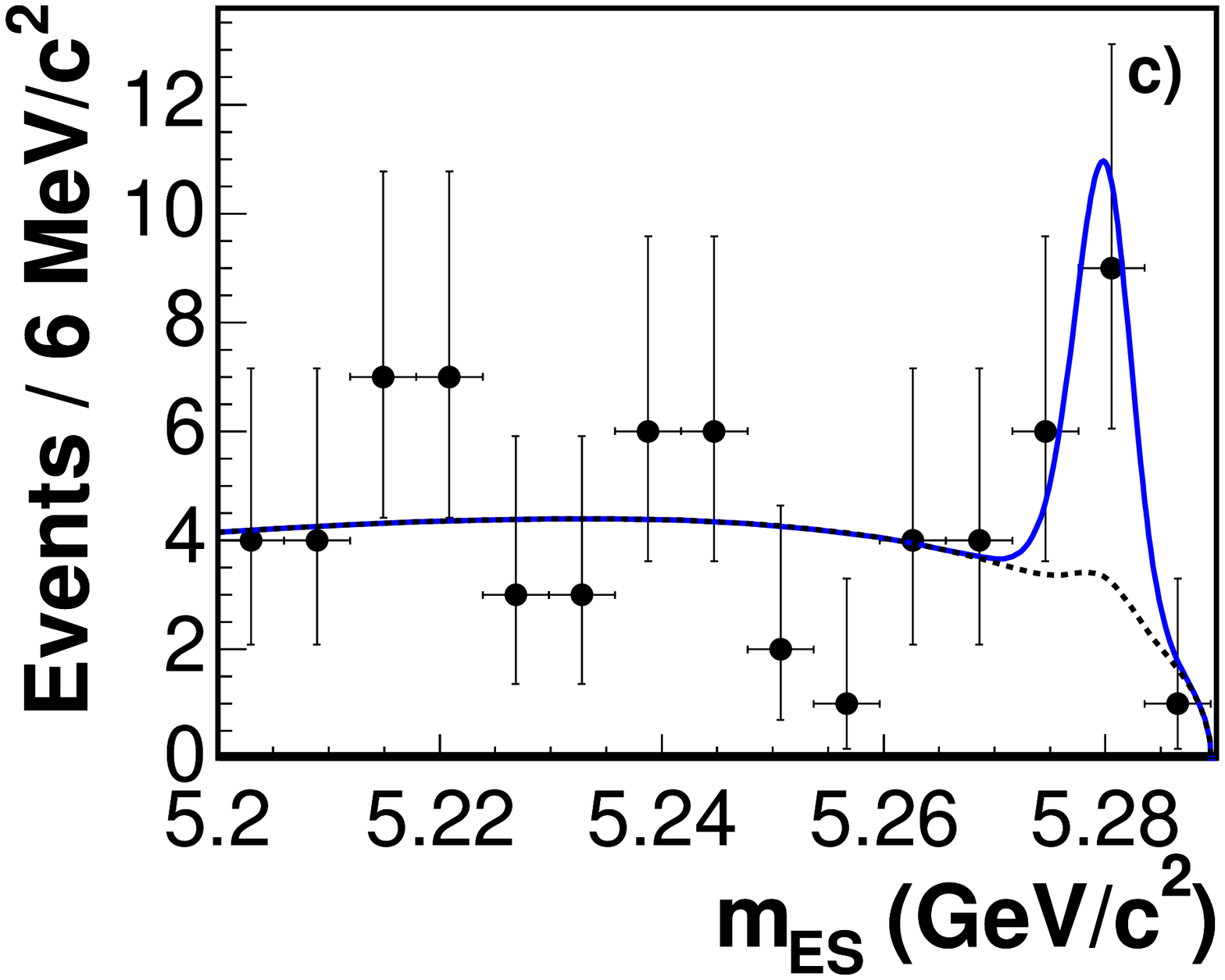}\includegraphics[width=0.50\linewidth]{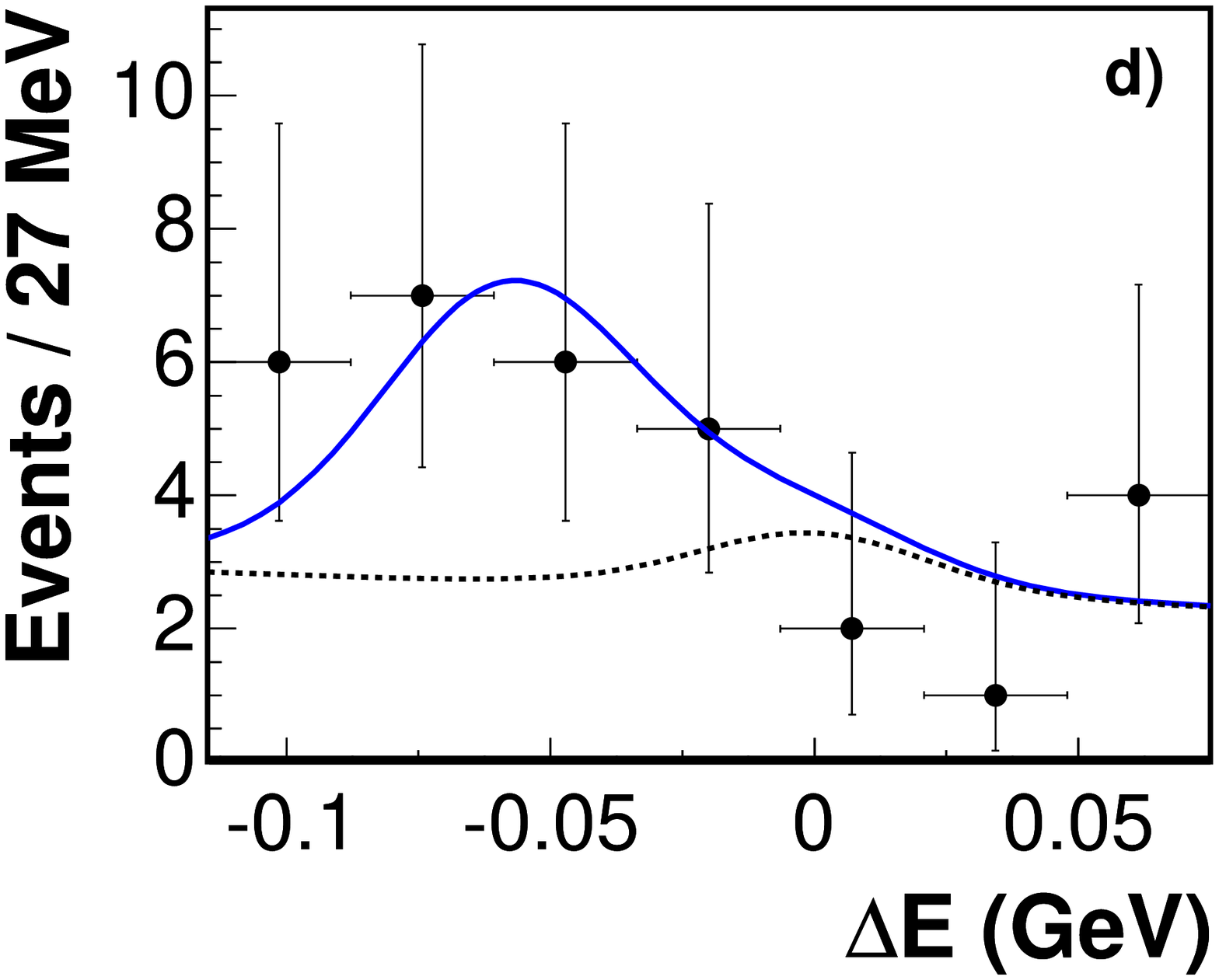}
\vskip1mm
\includegraphics[width=0.47\linewidth]{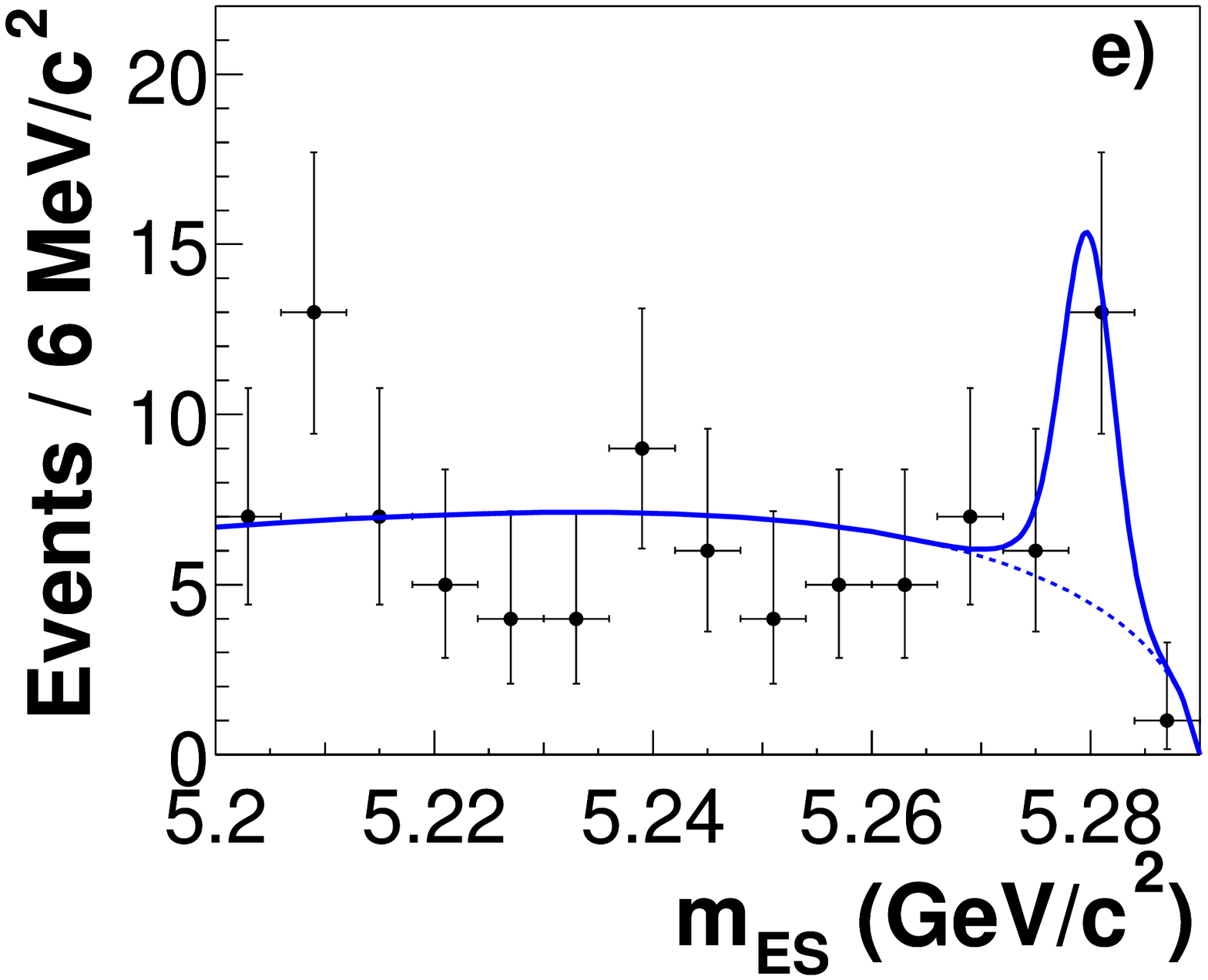}\hspace{0.2cm}\includegraphics[width=0.47\linewidth]{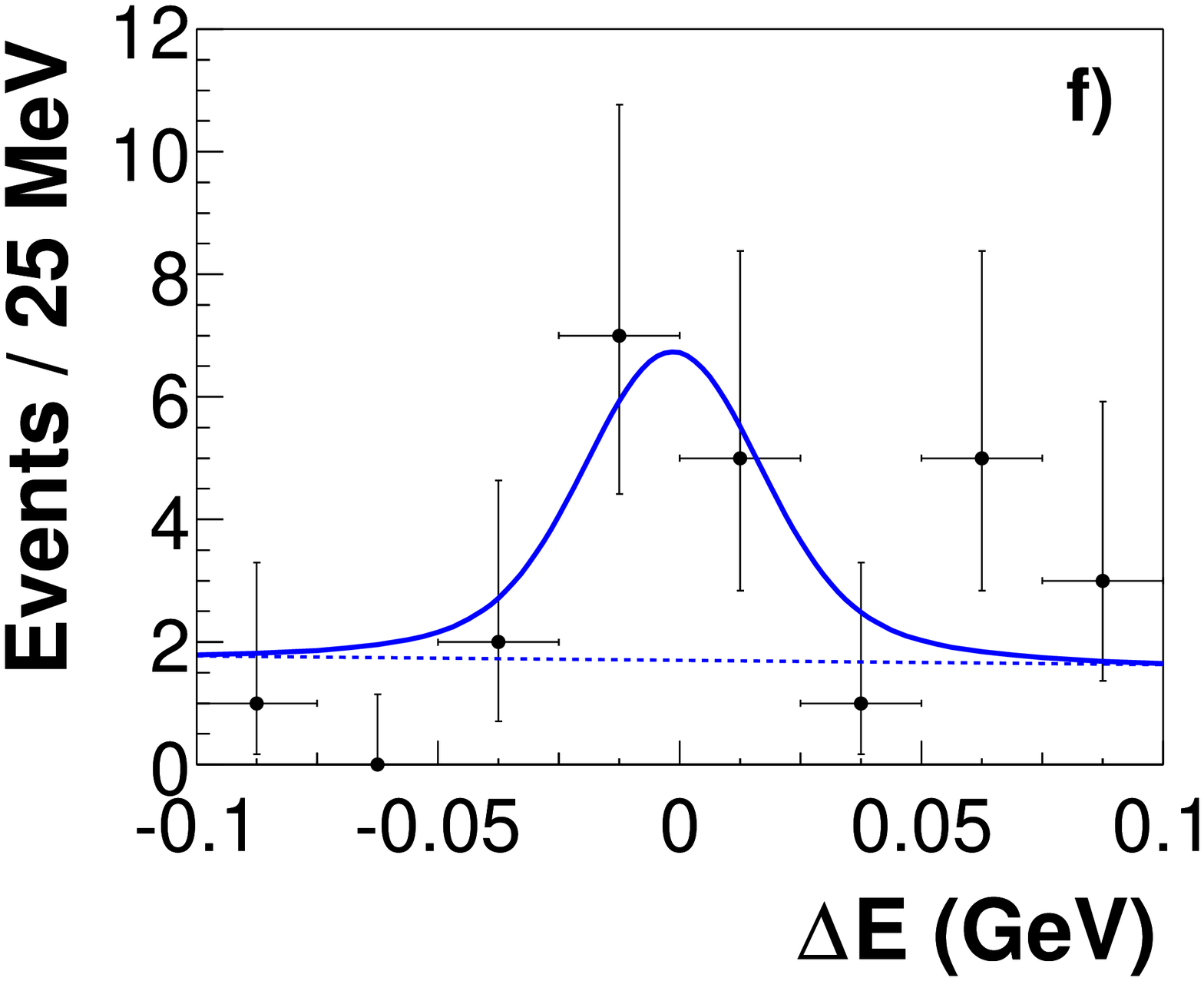}
\caption{Distributions of (a) $\mes$ and (b) $\de$ for signal (main plot) and 
background (inset) $\Bu\to\KS\pip$ candidates (points with error bars) 
using the weighting technique described in the text.  Solid curves represent 
the corresponding PDFs used in the fit.  In (c-f) we show projections of $\mes$ 
and $\de$ for $\KS\Kp$ (c,d) and $\KS\KS$ (e,f) decays (points with error bars) 
enhanced in signal decays using additional requirements on probability ratios.  
Solid curves represent the PDF projections for the sum of signal and background 
components, while the dotted curve shows the contribution from background only.}
\label{fig:sPlots}
\end{center}
\end{figure}

We compare data and PDFs in the high-statistics $\KS\pip$ mode using
the event-weighting technique described in Ref.~\cite{sPlots}.  For
the plots in Figs.~\ref{fig:sPlots}~(a,b), we perform a fit excluding 
the variable being shown; the covariance matrix and remaining PDFs are 
used to determine a weight that each event is either signal (main plot) 
or background (inset).  The resulting distributions (points with errors) 
are normalized to the appropriate yield and can be directly compared with 
the PDFs (solid curves) used in the fits.  We find good agreement between 
data and the assumed PDF shapes for $\mes$ and $\de$.  In 
Figs.~\ref{fig:sPlots}~(c-f), we show projections of the $\KS \Kp$ and $\KS\KS$
data obtained by selecting on probability ratios calculated from the 
signal and background PDFs (except the variable being plotted).  The 
solid curves in each plot show the fit result after correcting for the
efficiency of this additional selection.

Systematic uncertainties on the signal yields are due to the imperfect
knowledge of the PDF shapes.  We evaluate this uncertainty by varying the PDF 
parameters that are fixed in the fit within their statistical errors, and by 
substituting different functional forms for the PDF shapes.  For the charged 
modes, the largest contribution is due to the signal parameterizations for 
$\mes$ ($^{+13}_{-15}$ events for $\KS\pip$, $^{+1.3}_{-1.7}$ events for
$\KS\Kp$) and $\de$ ($^{+16}_{-5}$ events for $\KS\pip$, $^{+2.8}_{-0.7}$ events
for $\KS\Kp$), while for the neutral mode it is due to uncertainty in the 
background $\mes$ shape ($\pm 0.7$ events) and the potential fit bias 
($\pm 1.4$ events).  The systematic uncertainties on efficiency estimates are 
dominated by the selection on $\cos{\theta_S}$ $(2.5\%)$ and the uncertainty 
($1.2\%$ per $\KS$) in $\KS$ reconstruction efficiencies evaluated in a large 
inclusive sample of $\KS$ mesons reconstructed in data.  For the 
charge-asymmetry measurement, we use the background asymmetry to 
set the systematic uncertainty~\cite{KpiBaBar}.  We find background 
asymmetries of $-0.005\pm 0.010$ and $-0.002\pm 0.011$ for $\KS\pi$ and $\KS K$ 
events, respectively.  Both results are consistent with zero bias, and we 
assign the statistical uncertainty $(0.01)$ as the systematic error on 
${\cal A}_{\CP}(\KS\pip)$.

In summary, we find evidence for the decays $\Bu\to\Kzb\Kp$ and
$\Bz\to\Kz\Kzb$ with branching fractions on the order of $10^{-6}$ and 
significances of $3.5\sigma$ and $4.5\sigma$, respectively, including 
systematic uncertainties.  These results represent direct evidence for 
the $b\to dg$ penguin-decay process.  The branching fractions 
are consistent with recent theoretical estimates~\cite{theory}, implying 
that soft rescattering effects may not play an important role in these decays.  
We also measure the branching fraction 
$\BR(\Bu\to\KS\pip)=(26.0\pm 1.3\pm 1.0)\times 10^{-6}$
and the \CP-violating charge asymmetry 
${\cal A}_{\CP}(\KS\pip) = -0.09\pm 0.05\pm 0.01$, which 
are both consistent with previous measurements by other 
experiments~\cite{BelleAndCleo,BelleAndCleoACP}.

We are grateful for the excellent luminosity and machine conditions
provided by our \pep2\ colleagues, 
and for the substantial dedicated effort from
the computing organizations that support \babar.
The collaborating institutions wish to thank 
SLAC for its support and kind hospitality. 
This work is supported by
DOE
and NSF (USA),
NSERC (Canada),
IHEP (China),
CEA and
CNRS-IN2P3
(France),
BMBF and DFG
(Germany),
INFN (Italy),
FOM (The Netherlands),
NFR (Norway),
MIST (Russia), and
PPARC (United Kingdom). 
Individuals have received support from CONACyT (Mexico), A.~P.~Sloan Foundation, 
Research Corporation,
and Alexander von Humboldt Foundation.


\begin{thebibliography}{99}
\bibitem{KsKl}
The decay $\Bz\to\Kz\Kzb$ proceeds in an $s$ wave, which
produces equal fractions of $\KS\KS$ and $\KL\KL$, but no 
$\KS\KL$, neglecting \CP\ violation in the kaon system.
\bibitem{significance}
The effect of the systematic uncertainty on the significance estimate is 
evaluated in a conservative way by coherently shifting all fixed PDF 
parameters such that the yield is minimized.
\end{thebibliography}
\end{document}